\pdfoutput=1

\documentclass[letterpaper,12pt]{article}

\usepackage{arxiv}
\usepackage[T1]{fontenc}    
\usepackage{hyperref}       
\usepackage{url}            
\usepackage{booktabs}       
\usepackage{amsfonts}       
\usepackage{nicefrac}       
\usepackage{microtype}      
\usepackage{lipsum}
\usepackage{graphicx}
\usepackage[T1]{fontenc} 
\usepackage{graphicx,color}
\usepackage[dvipsnames]{xcolor}
\usepackage{soul}
\usepackage{url}
\usepackage{subcaption}
\usepackage{amssymb}
\usepackage{float}
\usepackage{pdf14}
\usepackage{tikz}
\usepackage{placeins}
\usepackage{tabularx}
\usepackage{schemata}
\usepackage{adjustbox}
\usepackage{graphicx}
\graphicspath{{img/}} 
\usepackage{bm}
\usepackage{listings}
\usepackage{blindtext}
\usepackage{amsmath}

\usepackage{color}

\graphicspath{ {./img/} }

\title{Analysis of dark matter halo structure formation in $N$-body simulations with Machine Learning}

\author{
  Jazhiel Chac\'on \\ 
  Centro de Investigaci\'on en Computaci\'on, Instituto Polit\'ecnico Nacional,\\
  07738, Ciudad de M\'exico, M\'exico \\
  Instituto de Ciencias F\'isicas, Universidad Nacional Aut\'onoma de M\'exico,\\
  62210, Cuernavaca, Morelos, M\'exico.\\
  \texttt{chaconl2021@cic.ipn.mx} \\
\And
  Isidro G\'omez-Vargas \\
  Instituto de Ciencias F\'isicas, Universidad Nacional Aut\'onoma de M\'exico,\\
  62210, Cuernavaca, Morelos, M\'exico.\\
  \texttt{igomez@icf.unam.mx} \\
\And
 Ricardo Menchaca M\'endez \\
 Centro de Investigaci\'on en Computaci\'on, Instituto Polit\'ecnico Nacional,\\
 07738, Ciudad de M\'exico, M\'exico.\\
 \texttt{ric@cic.ipn.mx}\\
 \And
  J. Alberto V\'azquez \\
  Instituto de Ciencias F\'isicas, Universidad Nacional Aut\'onoma de M\'exico,\\
  62210, Cuernavaca, Morelos, M\'exico.\\
  \texttt{javazquez@icf.unam.mx} } 
  
\begin{document}
\maketitle
\begin{abstract}
 The properties of the matter density field in the initial conditions have a decisive impact on the features of the large-scale structure of the Universe as observed today. These need to be studied via $N$-body simulations,  which are imperative to analyze high density collapsed regions into dark matter halos. In this paper, we train Machine Learning algorithms with information from $N$-body simulations to infer two properties: dark matter particle halo classification that leads to halo formation prediction with the characteristics of the matter density field traced back to the initial conditions, and dark matter halo formation by calculating the Halo Mass Function (HMF), which offers the number density of dark matter halos with a given threshold. We map the initial conditions of the density field into classification labels of dark matter halo structures. The Halo Mass Function of the simulations is calculated and reconstructed with theoretical methods as well as our trained algorithms. We test several Machine Learning techniques where we could find that the Random Forest and Neural Networks proved to be the better performing tools to classify dark matter particles in cosmological simulations. We also show that, by using only a few data points, we can effectively train the algorithms to reconstruct the Halo Mass Function in a model-independent way, giving us a highly accurate fitting function that aligns well with both simulation and theoretical results.

\end{abstract}


\section{Introduction}

\noindent
By studying the cosmological structure formation in the standard model, also known as $\Lambda$CDM, we are able to determine that the total amount of the Universe is divided among several constitutes: visible matter (baryonic matter),  which takes about 4.9 \% of the total amount; neutrinos and photons 
which today are estimated that from less than 0.1\% of the total content; dark matter, a hypothetical constituent of the Universe with purely gravitational interaction which collapses into filaments, halos and structures that eventually ended up merging the visible matter that creates galaxies and adds up about 23\% of the total content of the Universe and finally dark energy, another hypothetical constituent, but this one being the responsible for the current accelerated expansion of the Universe, with the remaining 72 \% of the total content of the Universe \cite{2013ApJS..208...20B}.  

The presence of dark matter and dark energy, the so called `dark sector' of the Universe, can be inferred from observational evidence of large-scale structures, which can be studied with analytical and semi-analytical models. Nevertheless, only numerical simulations are capable to emulate the small scale structures and sub-structures observed in the Universe, that give rise to a cosmic network of filaments, voids and groups. These structures act as gravitational wells around the visible matter, which eventually merge visible matter to create clusters of galaxies, quasars and gas clouds.

Using numerical simulations as virtual environments, we can evaluate and evolve a set of initial conditions of matter and energy that will eventually end up merging the structures as observed today by different telescopes and available probes. These environments can be considered as virtual laboratories that allow us to study different candidates of dark matter and dark energy. However,  until a few years ago, numerical simulations were very restrictive in terms of computational resources, and only accessible to a few research groups.  
In this regard, there has been a rising amount of alternative methods, hardware and algorithms that would allow to lower the computational resources required to carry a full run of a cosmological simulation. Using artificial intelligence and other machine learning methods is becoming more accepted to model relevant features in numerical simulations. Within this area of study, there has been a variety of tasks and hardware development to enhance the analysis of cosmological structure formation and data. Some of these methods range from  classification, clustering, regression, statistics and optimization among others \cite{gomez2023neural, gomez2023neuralepjc,2013ApJ...772..147X, 2015JCAP...01..038H, 2020arXiv200512276M, Kamdar_2016, Buncher_2020, 5290959,perraudin2019cosmological}.

The main idea presented in this work was described in \cite{CHACON2022100527} and originally based on the studies first made in  \cite{10.1093/mnras/sty1719}, however, this new approach was carried out on larger $N$-body simulations which increase the dataset with more dark matter particles and we tested them on a variety of machine learning algorithms. Hence we obtained improved results compared to those achieved in the previous work, also because the important statistics, such as the Matter Power Spectrum, which remain intact even with the change in scale. Implementing a wide variety of machine learning algorithms, that can be used for classification, helps to test their efficiency and viability in terms of computational cost, therefore we evaluated their classification performance and their computational runtime. We find out how much information the features in the initial conditions provide in order to determine the formation of dark matter halos in cosmological simulations. In addition, to explore another Machine Learning application in this cosmological field, we implement a neural network and a Gaussian Process to reconstruct, in a model-independent way, the Halo Mass Function, given only a few of points from simulations, which in most cases are computationally expensive to produce.

The content of this paper is as follows. In Section \ref{sect:2}, we describe briefly the ML classification methods. In the sections \ref{sect:3} and \ref{sect:4} we describe how we used our dataset as a binary classification problem. For this purpose, we present a dark matter particle halo formation framework, which uses information about local density field features in the initial conditions of the simulation, and the final dark matter halo formation. In Section \ref{sect:5} we present our discussion of the results achieved; in particular \ref{sec:results_classification} contains our classification results with several Machine Learning methods, and Section \ref{sec:results_nonparametric_HMF} includes the model-independent reconstructions for the Halo Mass Function. Finally, Section \ref{sec:conclusions} shows our final discussion and conclusions. 

\noindent

\section{Machine Learning algorithms}\label{sect:2}
\noindent
Machine Learning (ML) is the field of Artificial Intelligence that is focused on the statistical modeling of data. Common tasks in ML are classification, clustering, pattern recognition and time series analysis, among others. In this paper, we use ML algorithms to perform the classification of particles from N-body simulations in order to know whether they are within a dark matter halo, or they are not. Then, we briefly introduce the algorithms used in this work.

\subsection{Logistic regression}
\noindent
 Logistic regression, or logit regression, is a linear model for classification. It is one of the pioneer classification algorithms in Machine Learning, which relies its mechanism on assigning, to each instance, a probability of belonging to a particular class using the logistic function (also called sigmoid): 

 \begin{equation}
     \sigma(x) = \frac{1}{1+e^{-x}}.
 \end{equation}
 
 This algorithm works in the same way as a linear regression, with the difference of converting its outputs into probabilities using the logistic function. We can describe the logistic regression as follows:
 
 \begin{equation}
     \hat{p} = \sigma(\theta^T \cdot x),
 \end{equation}
 where $\theta$ represents the straight-line parameters and $x$ the features of the input data. Therefore, the classification $y_{\rm pred}$ is given by:
 \begin{equation}
    \schema
	{
	\schemabox{$y_{\rm pred}$}
	}
	{
	\schemabox{0 \; if  \; $\hat{p}$ $<$ 0.5 \\ 1 \; if  \; $\hat{p}$ $\geq$ 0.5 }
	},
 \end{equation}
 where $0$ and $1$ are two different classes.
 
\subsection{Bayes classifier}
\noindent
A Bayes classifier obtains the conditional probability of each class $C_i$ given a set of of $n$ attributes $A = {A_1, ..., A_n}$ through the Bayes rule:
\begin{equation}
    P(C_i|A) = \frac{P(C_i)P(A|C_i)}{P(A)}, 
\end{equation}
where the associated parameters are the prior probability for each class $P(C_i)$ and the conditional probability for each attribute given a class $P(A|C_i)$; these parameters can be estimated from the dataset using frequencies, i. e., this is a frequentist approach of the Bayesian rule.

\noindent
The simplest Bayes classifier, known as Naive-Bayes, assumes that the attributes are conditionally independent; however, in several datasets, this condition is not satisfied and for this reason, some proposals have emerged that are its extensions, such as the Naive-Bayes complement \cite{rennie2003tackling}, Gaussian Naive-Bayes \cite{chan1982updating} and multimodal Naive-Bayes \cite{manning2008introduction}. In this work, we use these mentioned extensions and the classical Naive-Bayes algorithm.

\subsection{Support Vector Machines}
\noindent
 A Support Vector Machine (SVM) is a supervised learning method that uses separating hyperplanes in high-dimensional spaces to perform classification. It is particularly useful for solving problems where the number of features exceeds the number of observations, and the data points are not easily separable.
 
 The primary concept behind SVM is to identify the hyperplane that separates the data points of distinct classes to the maximum extent possible. This hyperplane is chosen in a way that maximizes the margin, which refers to the distance between the hyperplane and the nearest data points of each class. The support vectors are the data points that are closest to the hyperplane and are used to define the hyperplane. The objective of SVM is to determine the optimal boundary between the potential classes. While there are several hyperplanes that could potentially separate the classes, the best option is the one that has the greatest distance between the points of different classes, known as the maximum-margin hyperplane. Therefore, SVM involves a maximization problem.
 
 SVMs have the ability to handle high-dimensional data and 
 they are robustness to overfitting. A concise explanation about SVM is available in Ref. \cite{noble2006support} and for mathematical details we recommend the Ref. \cite{smola2004tutorial}.

\subsection{Decision trees and random forest}
\noindent
Decision trees refer to a paradigm of learning based on approximating discrete target functions, in which the learned function is represented by a decision tree \cite{mitchell1997machine}. The elements of a decision tree are the roots (where the data is stored), the branches (the path the tree takes to make decisions) and the nodes (consisting of sets of elements that have a determined characteristic after a decision is made). Given a dataset, we can calculate the inconsistency within the set, or in other words, find its entropy in order to divide or split the set until all data are within a given class \cite{quinlan1986induction}. When there exist a large number of decision trees that operate together as an ensemble, we are referring to the Random Forest algorithm \cite{hastie_2009}. The randomness of this algorithm comes from the fact that operations and predictions from the forest are not hierarchically taken, but a subset of elements (like the number of trees, number of attributes, length of data, etc.) is taken in a random way.

\subsection{Artificial Neural Networks}
\noindent
Artificial neural networks (ANNs) are able to model large and complex datasets and any nonlinear function \cite{hornik1990universal}. They are computational models that represent the synapse of biological neurons through interconnected layers of units called neurons or nodes, which make up its basic information processing elements. In the simplest type of neural network, the  Multi-layer perceptron (MLP) (also called feedforward neural network), there are three types of layers: an input layer that receives the input, hidden layers responsible for extracting patterns and producing nonlinearity effects, and finally the output layer that presents the results of the prediction. 

For a full background of neural networks we recommend the references \cite{goodfellow2016deep, bishop2006pattern, nielsen2015neural}; or, for a basic introduction in the  cosmological context, see \cite{rojas2022observational}.

\subsection{Gaussian Processes}
The Gaussian Processes (GPs) algorithm \cite{mackay2003information, dudley_2002} is a powerful probabilistic approach used in machine learning to model the relationship between inputs and outputs. It offers several advantages, including the ability to handle noisy or incomplete data and provide uncertainty estimates for predictions. GPs define a prior distribution over functions, where any finite set of function values follows a joint Gaussian distribution. Given a set of observed input-output pairs, the goal is to infer the underlying function that best explains the data and quantify the uncertainty in predictions.

Mathematically, GPs are characterized by a mean function and a covariance function (kernel). The mean function represents the expected value of the process at any input location, while the covariance function captures the similarity between inputs. By incorporating observed data, GPs compute a posterior distribution over functions. Predictions for new inputs can then be made by calculating the predictive mean and variance. Gaussian Processes are widely used in machine learning, especially in applications involving small data sets, because they can handle noisy or incomplete data and provides a measure of uncertainty in their predictions.

\section{Dark matter halo formation as a binary classification framework} \label{sect:3}

\noindent
Using the set of numerical simulations described in \cite{article-Chacon}, we are able to obtain a relational database that includes information about the merged dark matter halos at cosmological time $z = 0$. We obtained numerical features within the initial conditions of the dark matter particles used for the simulations. Additionally, we identified host halos and sub-halos that allow us to associate dark matter substructures to larger merged halos, in order to determine a dark matter halo mass classification threshold. 

%

\subsection{Data selection}\label{sec:3.1}
\noindent
We performed a simulation with the cosmological code GADGET-2, \cite{2005MNRAS.364.1105S} assuming a $ \Lambda$CDM Universe. The data output was designed such that the dimensionless density parameters were $\Omega_{m}$ = 0.268, $\Omega_{\Lambda}$ = 0.683, $\Omega_{b}$ = 0.049, $h$ = 0.7 and with a gravitational softening of $ \epsilon $ = 0.89 kpc. The total number of dark matter particles is $ 192^{3} $, each with a mass of 1.3 $\times 10^{9} $ M$_{\odot} $ in a box of comoving length $ L = 50 h^{- 1} $ Mpc running from $ z = 23 $ to $ z = 0$. To identify dark matter halos and subhalos we use ROCKSTAR halo finder \cite{2013ApJ...762..109B}. The final snapshot has a total of 4000 dark matter halos whose masses fall within the range $(10^{11} \leq M/M_{\odot} \leq 10^{14})$. This given mass will be used as the binary threshold to classify dark matter particles: if a selected particle falls within a halo with mass between $10^{11} \leq M/M_{\odot} \leq 10^{14}$, then the dark matter particle will have a classification label of \texttt{1}, otherwise if this condition is not met, the dark matter particle will have a classification label of \texttt{0}.

The properties from the density field of the initial conditions of the simulation serve as an input data to our ML models. The attributes of the dataset are calculated from analytical works related to the halo mass function (HMF) by Press-Schechter \cite{1974ApJ...187..425P}. This function predicts the number density  of dark matter halos depending on their mass and the density field. The density generates a halo of a certain mass $ M $ at a given redshift $ z $. If it exceeds a critical value $ \delta_{c} (z) $, these values will be called overdensities at redshift $ z $.

The core idea is that the dark matter halos enclose their mass in a dense spherical region, where the density contrast will be given by the following relation:
%
\begin{equation}
    \delta(\textbf{x}) = \frac{\rho(\textbf{x}) - \Bar{\rho}}{\Bar{\rho}},\label{eqn:4.1} 
\end{equation}
where $\Bar{\rho}$ is the average matter density of the Universe. For a sphere of radius $R$, the overdensity is defined as follows \cite{dodelson2003modern}:
\begin{equation}
    \delta(\textbf{x},R) \equiv \int d^{3}\textbf{x}'\delta(\textbf{x}')W_{R}(\textbf{x} - \textbf{x}').\label{eqn:4.2}
\end{equation}
In eq. (\ref{eqn:4.2}), $W_{R}$ is a window function of the \textit{top hat} model, given by:
\begin{equation}
W_{R} = \left\lbrace
\begin{array}{ll}
\frac{3}{4\pi R^{3}} & \textup{if} \; |\textbf{x}| \leq R\\
0 & \textup{if} \; |\textbf{x}| > R.
\end{array}
\right.\label{eqn:4.3}
\end{equation}
A window function with radius $ R $ and volume $V$ corresponds to a mass scale $M = \Bar{\rho} V(R)$. The expected value of the overdensity in eq. (\ref{eqn:4.2}) is the normalization term of the power spectrum $ \sigma_{R} $, which is defined by the following relation:
\begin{equation}
    \sigma^{2}_{R} = \langle \delta^{2}(\textbf{x}, R)\rangle.\label{eqn:4.4}
\end{equation}
The attributes that we are able to define with eqs. (\ref{eqn:4.2}) and (\ref{eqn:4.3}) allow us to create a structured database, with the information of overdensities at different radius $R$ values, derived from a mass scale $M_{R}$ centered on the position of a particle, from the initial conditions at redshift $z = 23$. This allows us to create a ten-component vector, namely, $\delta_{1},...\delta_{10}$,  with their respective classification label. The data we selected takes into accout two main features; first, the data that we are occupying makes use of the average mass range of the halos found; second, the
range of mass was also a calculation from the spherical collapse model, which gave us the range of $(10^{11} \leq M/M_{\odot} \leq 10^{14})$ for the threshold.

\begin{figure*}[t!]
    \centering
    \includegraphics[width = \textwidth]{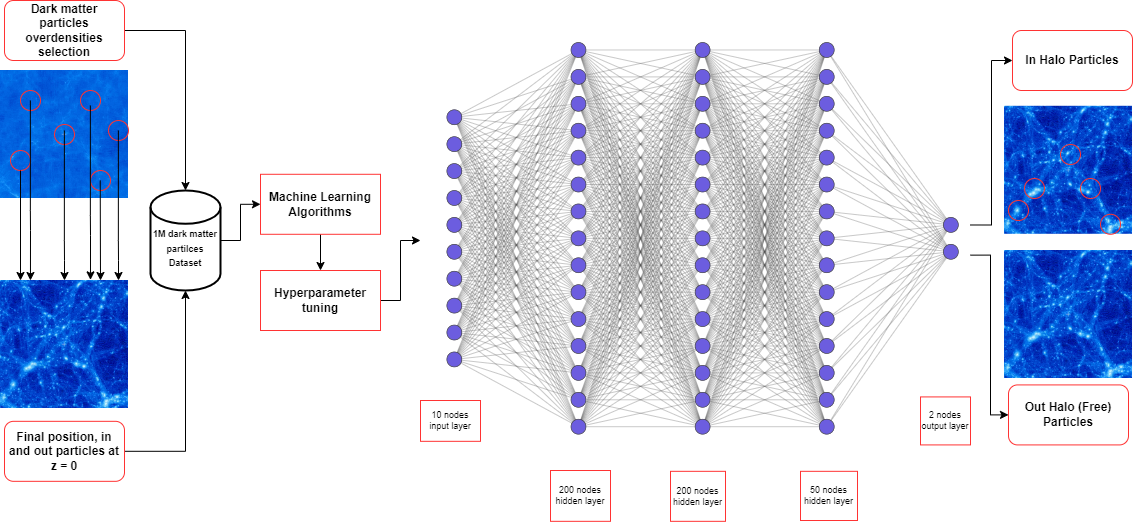}
    \caption{ 
    \footnotesize{Diagram of the method to select the properties of the initial density field conditions that will eventually form the structure in the simulation. The process starts by extracting properties of the initial conditions in the local neighborhood of the density field around dark matter particles and associate them with the final position in the halo distribution. The final classification \texttt{Not in halo}, \texttt{In Halo} depends on the mass threshold chosen to determine whether a dark matter particle will belong to a halo or if it is not bound to any other object.}} \label{fig:1}
\end{figure*}


\section{Methodology} \label{sect:4}

\noindent
 The framework described in the previous section allowed us to implement the ML algorithms described in Section \ref{sect:2}, all of them available in the \texttt{Python} libraries: \texttt{Scikit-Learn} \cite{scikit-learn} and \texttt{TensorFlow} \cite{tensorflow2015-whitepaper}. The dataset consists of one million (1,000,000) randomly selected dark matter particles, each of them with a ten-component vector, whose features are the overdensities $\delta_{1},...\delta_{10}$ at different values of radius $R$. This selection upscales the previous work \cite{CHACON2022100527}, where we used 50,000 randomly selected particles, up to 2 orders of magnitude, allowing our models to be more competitive with the training and validation tests. Since the particles are randomly sampled from the simulation, it is unlikely they are in some way correlated, this allows us to reduce the bias and overfitting at the moment of evaluation in the test sets. Figure \ref{fig:1} shows a schematic picture for our entire pipeline to train our algorithms.

All classification algorithms were fine-tuned by performing a hyperparameter grid search. This grid has a variety of hyperparameters depending on the algorithm we were testing. Each algorithm will have as an outcome the accuracy, precision, recall and F$_{1}$-score as their performance metrics. The algorithms were successfully trained and we tested their ability to predict the final label of the particles in the test set, which is compared with the real labels in order to obtain the performance of each algorithm. This evaluation was carried out under the Receiving Operation Characteristics (ROC) curve \cite{FawcettROC} along with the Area Under Curve (AUC).

\section{Results}\label{sect:5}
\subsection{Dark matter particles classification}
\label{sec:results_classification}

\begin{figure}[t]
    \centering
    \includegraphics[width = 0.6\textwidth]{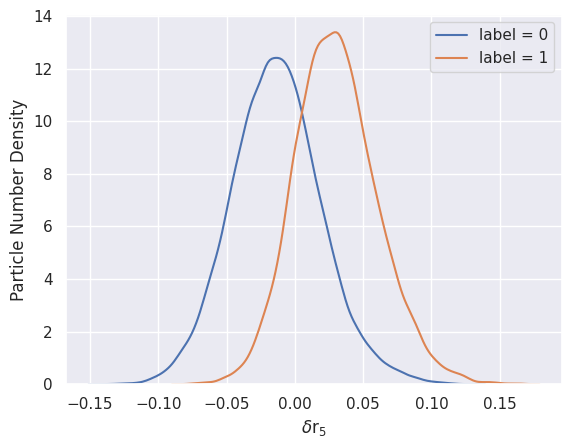}
    \caption{
    \footnotesize{Dark matter particle number density distribution according to a defined overdensity $\delta_{5}$. It can be noticed that our dataset is evenly distributed, we have more particles in the \texttt{In Halo} class rather than the \texttt{Not In Halo} class. Given this, we can use both ROC and PR curves to test the accuracy of the algorithms.}}
    \label{fig:2}
\end{figure}

\begin{figure}
    \centering
    \includegraphics[width = 0.7\textwidth]{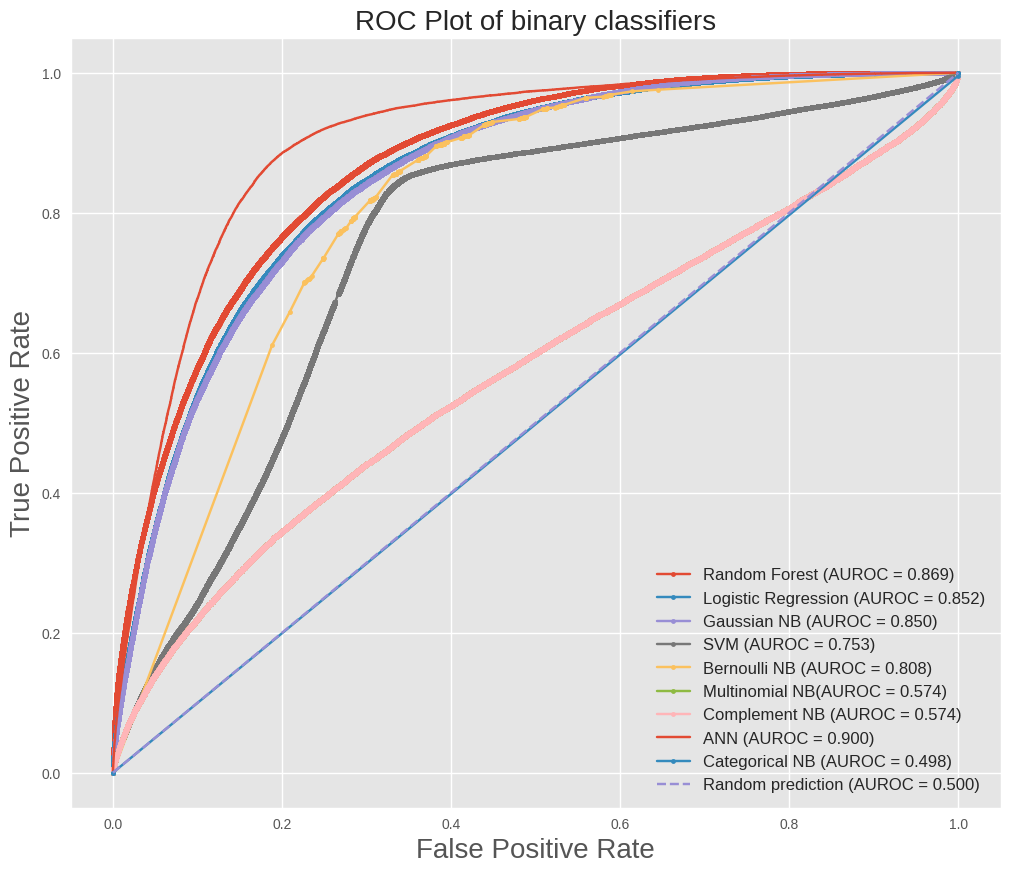}
    \caption{\footnotesize{ROC and PR curves of the 9 ML classifiers used. It can be seen that the performance rate drops drastically, especially for the Naive Bayes-like classifiers. This is related to the case that some of these classifiers perform better with categorical data rather than numerical data. These classifiers may not be used as predictors.}}
    \label{fig:3}
\end{figure}

\begin{figure*}[t]
    \centering
    \includegraphics[width = \textwidth]{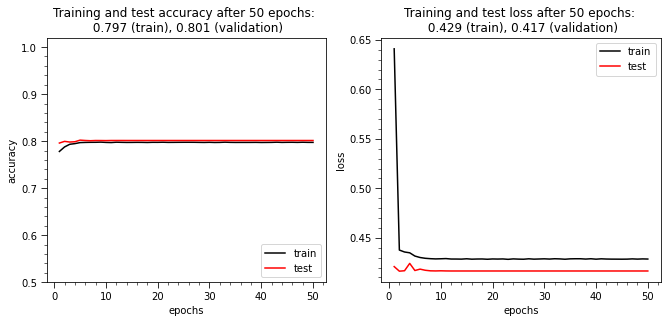}
\caption{\textit{Left}. Accuracy plot for the training  and test sets for the Multi-Layer Perceptron through 50 epochs. The low gap between training and test sets is an indicator of how well is performing our neural network. \textit{Right}. Loss function behavior on the training and test sets of the Multi-Layer Perceptron.}
    \label{fig:4}
\end{figure*}


\begin{table*}%
\caption{Metrics and scores of binary classifiers}
\begin{tabularx}{\textwidth}{@{\extracolsep{\fill}}l c c c c }
\hline%
Classifier & Accuracy & Precision & Recall & F$_{1}$ score \\\hline 

Logistic Regression     & 0.722 & 0.498 & 0.711 & 0.727 \\
Naive Bayes Gaussian    & 0.718 & 0.516 & 0.757 & 0.748 \\
Naive Bayes Bernoulli   & 0.699 & 0.495 & 0.757 & 0.730 \\
Naive Bayes Multinomial & 0.611 & 0.316 & 0.554 & 0.527 \\ 
Naive Bayes Complement  & 0.608 & 0.316 & 0.554 & 0.527 \\
Naive Bayes Categorical & 0.699 & 0.370 & 0.578 & 0.568 \\ 
Support Vector Machines & 0.723 & 0.473 & 0.753 & 0.702 \\ 
Random Forest           & 0.785 & 0.767 & 0.792 & 0.792 \\
Multi Layer Perceptron  & 0.806 & 0.764 & 0.794 & 0.794 \\
\hline%
\end{tabularx} \label{Table:1}
\end{table*}

\noindent
We were able to test nine binary classification algorithms using the $N$-body simulation data as input. In figure \ref{fig:2} we plot the dark matter particle distribution labels, where we observe how our selection is evenly distributed. This is an indicator to verify the results using the ROC curve and the AUC score, resulting in the performance plotted in figure \ref{fig:3}. Additionally, we detail the performance metrics in Table \ref{Table:1}. It is interesting to point out that the best performing classifier was the Multi-Layer Perceptron, which obtained a high accuracy of 80\%, and that its closest contender was the random forest, with 78\% accuracy.

On the other hand, logistic regression has good performance in terms of accuracy and execution time, this is due to the own algorithm's design, which is performing quick calculations with the sigmoid function, but it lacks precision, which means it is predicting a high amount of false positives. Support vector machines are another example where having a high accuracy does not imply having high precision, this low percentage is giving rise to almost 50\% data as false positives. These results are not optimal for prediction purposes.

Regarding other classifiers, even though they have a good performance overall, they fail to perform better on unseen data due to the low precision and recall, as it turns out, some of these classifiers are very well adapted to categorical data, like Naive Bayes Categorical or Multinomial, and since we used only numerical data, their precision is lower, this is an indicator to discard these types of algorithms as predictors in the test set.

An interesting result we obtained was with the neural network performance, in which we see the highest accuracy among all other algorithms tested. For this particular method, we use three hidden layers with 200, 200 and 50 nodes respectively, 50 epochs, 1024 as batch size, 0.2 for dropout value, sigmoid as activation function and with the Adam gradient descent algorithm. With this configuration, figure \ref{fig:4} shows a good performance for the neural network in the accuracy metric and in the loss function for the training and test sets. Furthermore, the Multi-Layer Perceptron obtained the highest metric values among all other classifiers, with a fairly low execution time. As we can observe from figure \ref{fig:4}, the training and test accuracy curves after 50 epochs are practically the same, which means that the neural network is correctly classifying unseen data and, additionally, the training and test loss values are diminishing after less than 10 epochs. This result is an indicator that our model is not overfitting. 

\subsection{Halo Mass Function reconstruction}
\label{sec:results_nonparametric_HMF}
We were also able to reproduce some interesting results regarding the Halo Mass Function (HMF) at redshift $z=0$ with our different algorithms. As we know, the halo mass function is a very important tool for understanding the large-scale structure of the Universe and the eventual formation of galaxies. It describes the statistical distribution of the masses of the dark matter halos and provides information of mass accretion  at different cosmological redshifts. It is a fundamental ingredient in both cosmological simulations and theoretical models of structure formation. It allows us to test and refine the underlying nature of the dark matter and the laws of physics governing our Universe. In this work, the halo number density in function of its mass is one of the main results that we want to focus when we evaluate a simulation output. The density profile has been widely studied since 1974, and one of the most important density profiles is the Press-Schechter profile \cite{1974ApJ...187..425P}, which was used in this work as a benchmark. 



Since halo formation is hierarchical, the less massive halos merge to create the most massive halos. The HMF is often displayed in $\log_{10}$ scale. This profile can be solved analytically  using some parametric modifications of the top hat spherical collapse  model. With this information, we use the parametric results of the Press-Schechter density profile with HMFcalc \cite{2013A&C.....3...23M}, an online tool to obtain the HMF with the Press-Schechter formalism, with initial conditions similar to the $\Lambda$CDM scenario of our simulation, as described in section 3.1.    

Upsampling data with machine learning algorithms is an important technique that has a significant impact on the performance of many machine learning models and prediction. Increasing the number of samples in a dataset where the number of samples is significantly smaller than the prediction we want to achieve. Using various techniques like data augmentation from the MLP we can, therefore, balance the distribution and prevent the model from being biased towards the objective function, in this case, the halo mass function. This results in improved accuracy and precision in predicting and further generalization of unobserved data.


\subsection{Predicting the HMF with Gaussian Processes and Artificial Neural Networks}
In this particular scenario, we aimed to train a Gaussian Process algorithm in order to upscale and upsample the training dataset that we obtained directly from our simulation. Due to the scarcity and limitation of observations, it is a challenge to find the best hyperparameters needed to reconstruct the HMF without losing relevant physical information. Our dataset is passed through the GP algorithm in 50 iterations with the training data. The results are observed in figure \ref{fig:5_HMF}.

\begin{figure}[t!]
    \centering
\begin{minipage}[b]{0.49\textwidth}
    \includegraphics[height=5.2cm, width=7.8cm]{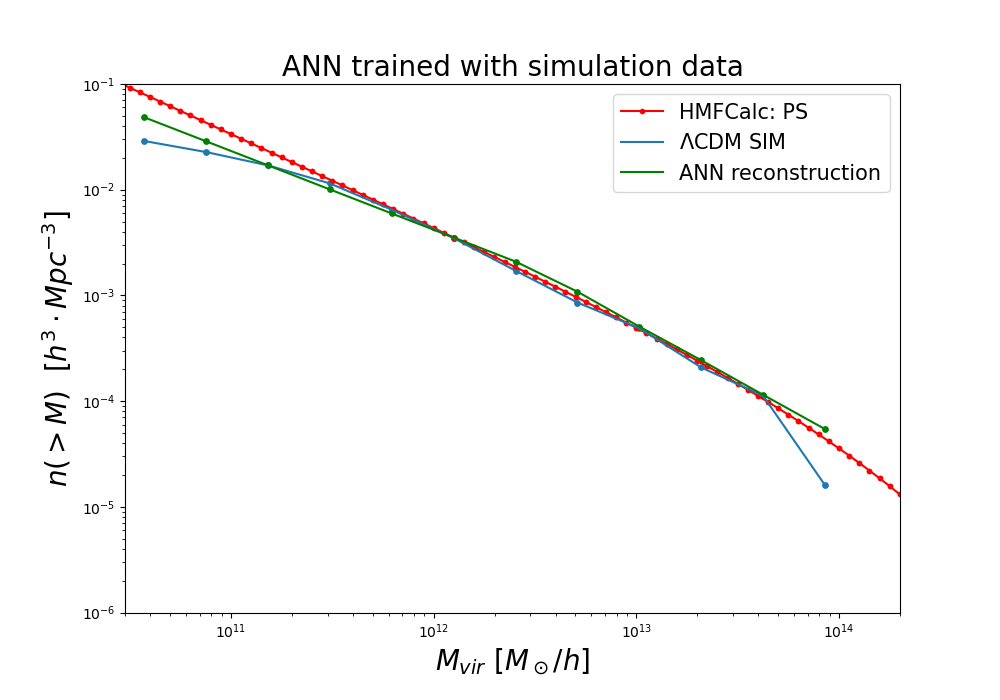}
 \end{minipage}
  \begin{minipage}[b]{0.49\textwidth}
    \includegraphics[height=5.2cm, width=7.8cm]{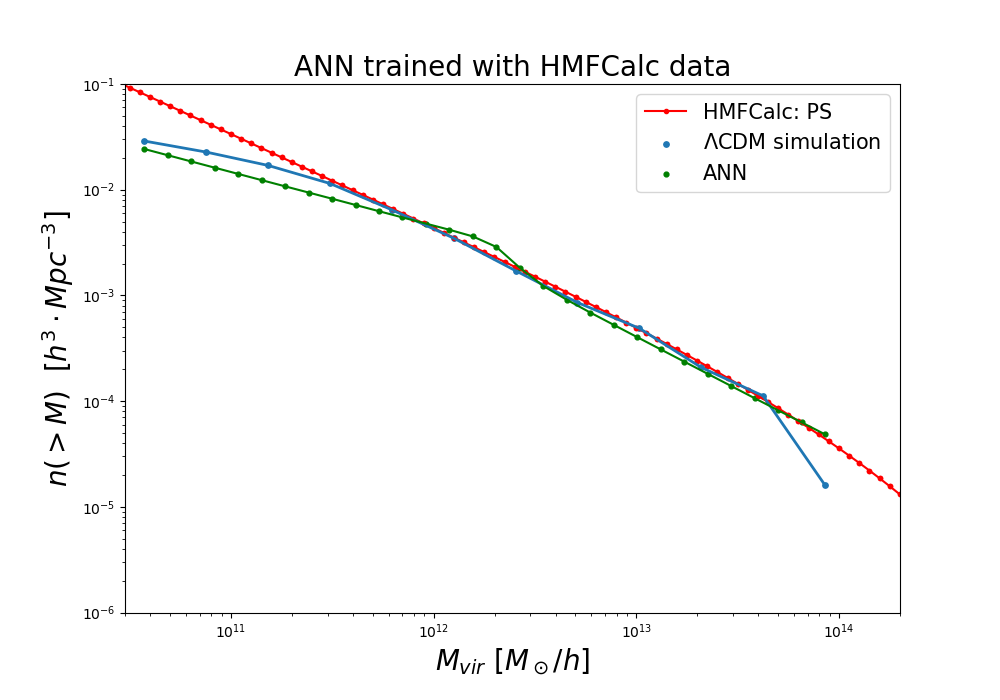}
  \end{minipage}
  \caption{\textit{Left}: Halo Mass Function reconstructed by the ANN (green line) trained and fitted with the $\Lambda$CDM simulation data (blue line). Since the ANN does not learn very well from the extreme values \cite{2021arXiv210400595G}, it is visible that the ANN does not follow the pattern on values higher than $10^{13} M_{\odot}$ and fewer than $10^{11} M_{\odot}$. \textit{Right}: Halo Mass Function reconstructed by the ANN (green line) trained and fitted with HMFCalc data (red line). We were able to upsample the initial datapoints available with this treatment.} \label{figure5}
\end{figure}

\begin{figure}[t!]
    \centering
    \includegraphics[width =0.7\textwidth]{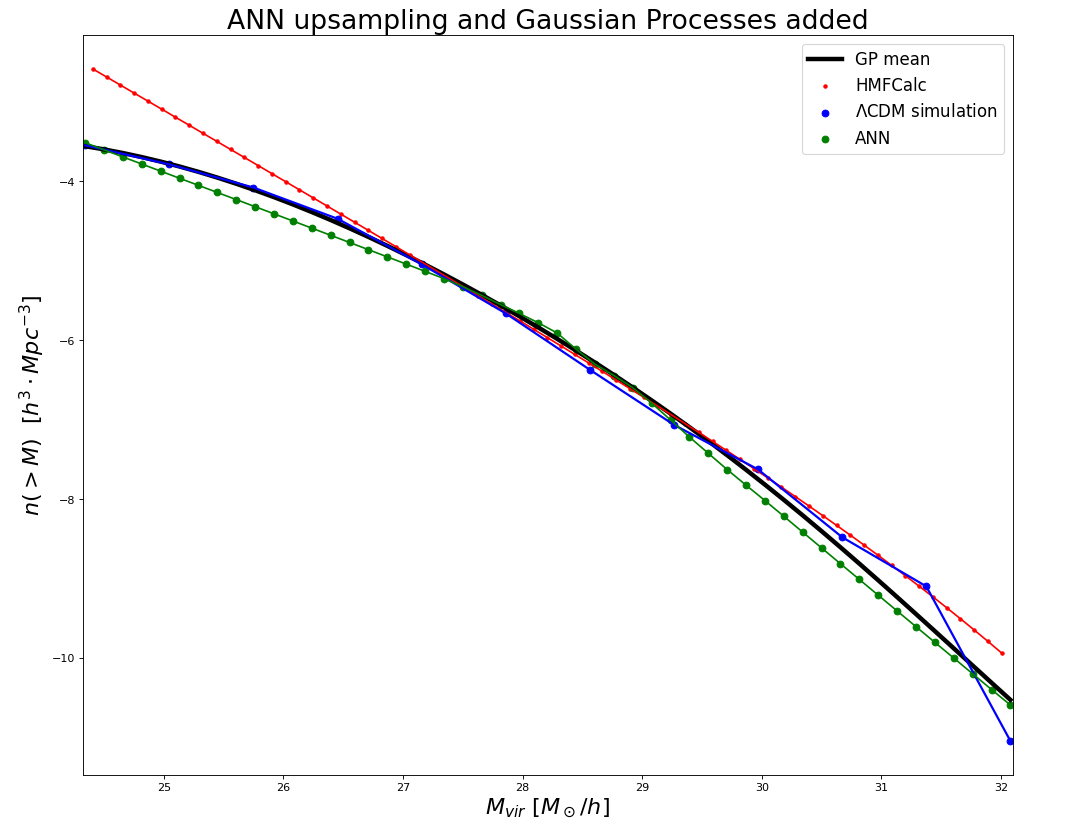}
    \caption{Halo mass function reconstruction with the ANN (green line), Gaussian Processes (dark line), compared with both simulation data (blue line) and HMFCalc with a Press-Schechter profile (red line). As observed,  we obtained few data from the simulation, however, with our implementation, it is possible to obtain more information with the Neural Network. The NN performs specially well in the $[10^{12},10^{13}M_{\odot}]$ range, whereas in the more extreme values, it tends to extrapolate the result with the simulated data. GPs also perform well and generate a good model even at extreme values.}
    \label{fig:5_HMF}
\end{figure}

Using only 13 data points from cosmological simulations, we were able to reconstruct the HMF with an ANN, being careful with the selection of its hyperparameters \cite{gomez2023neuralepjc}. Figure \ref{figure5} shows the result of using only the ANN, where we can see how the reconstruction fits the data of the simulation as well as the theoretical curve of HMFcalc. On the left side, we observe the HMF reconstructed with the training and fitting data from the $\Lambda$CDM simulation, whereas, on the right, we upsampled the data available using the HMFCalc values from the Press-Schechter formalism. In figure \ref{fig:5_HMF} we can see the result of using simulation and HMFCalc as training data for our ANN and GP methods. Particularly, the GP process was done in 50 iterations. Notice how the low sample points retrieved from the $\Lambda$CDM simulation are upsampled with the use of the MLP and how the GP fits with a continuous line, because these algorithms are extrapolating data in order to achieve better results, fitting the HMF. Since we are using a 50 Mpc boxsize, halo formation in the higher mass scale is somehow affected by this resolution. By using these two methods, we can predict higher mass halo merging, while in the lower bound, less massive halos are still within the range of the simulation, this may be due to the neural network does not learn very well from extreme values \cite{2021arXiv210400595G}. Nevertheless, the prediction of both algorithms fits in the $[10^{12}M_{\odot},10^{13}M_{\odot}]$ halo mass range. This result may seem contradictory, because it is known that in order to have a reliable result in the performance of an algorithm it is required to have a big amount of data, however, the performance of both algorithms showed to be good in the higher bound mass values. The Multi-Layer Perceptron and the Gaussian Process were trained in the middle range values and their predictions fit the results of both theoretical and simulated data.  

\newpage

\section{Final discussion and future work}
\label{sec:conclusions}
We were able to improve the results found in Reference \cite{CHACON2022100527} by implementing more classification algorithms to an enlarged data selected from a $N$-body cosmological simulation, our emphasis was to show a selected number of ML classifiers and see how well the performance showed on unseen data. As it turned out, some machine learning methods outperformed, like random forest and neural networks, since they rely more on iterative processes over numerical data to train and test their predictions, whereas categorical algorithms do not fit our binary classification purposes.
We can conclude that our dataset, consisting of feature selection of the initial conditions of the dark matter density field together with the final halo formation, has enough information to provide insight into the algorithms used as well as the physical properties within the data selected. The training process was refined up until it gets the best hyperparameters, and  since the number of selected particles represents a bigger percentage of the total number of particles within the simulation, we observed an improvement in performance compared to our previous results.

On the other hand, we have achieved significant findings through our employment of ANNs and Gaussian Processes for model-independent reconstructions. Our approach presents an alternative perspective on the formation of halo number density in cosmological simulations, utilizing a relatively small dataset. Remarkably, our results exhibit exceptional accuracy within the halo mass range of $[10^{12}M_{\odot},10^{13}M_{\odot}]$ halo mass range, obviating the necessity for a large volume of training data. Both methods demonstrate excellent agreement with the $\Lambda$CDM simulation, suggesting their efficacy as computational models for the Halo Mass Function. Furthermore, these models enable rapid interpolations based on sparse data points, eliminating the need for additional simulations.

We are encouraged to continue this work with advanced deep learning techniques using image information from $N$-body simulations in the near future to enrich our classification and regression applications.

\section*{Acknowledgments}
\noindent
J.Ch. would like to thank CONACYT for providing scholarship support and Brian Daniel Manzano Quechol for his help in the process of this work.
IGV thanks the CONACYT postdoctoral fellowship and the support of the ICF-UNAM. J.A.V. acknowledges support from FOSEC SEP-CONACYT Ciencia B\'asica A1-S-21925, FORDECYT-PRONACES-CONACYT 304001 and UNAM-DGAPA-PAPIIT IN117723. 

\bibliographystyle{abbrv}
\bibliography{main.bib}

\end{document}